\RequirePackage{fix-cm}
\documentclass[twocolumn]{svjour3}          % twocolumn
\smartqed  % flush right qed marks, e.g. at end of proof
\usepackage{svg}
\usepackage{graphicx}
\usepackage{amsmath}
\usepackage{subcaption}
\usepackage{color}
\usepackage[numbers]{natbib}
\usepackage{glossaries}

\usepackage{pgfplots}

\usepackage{tikz}
\usetikzlibrary{arrows}
\pgfplotsset{compat=newest}
\usepgfplotslibrary{fillbetween}
\usetikzlibrary{math}
\usetikzlibrary{calc}
\def\centerarc[#1](#2)(#3:#4:#5)% Syntax: [draw options] (center) (initial angle:final angle:radius)
{ \draw[#1] ($(#2)+({#5*cos(#3)},{#5*sin(#3)})$) arc (#3:#4:#5); }
\usetikzlibrary{shapes.geometric}
\usetikzlibrary{positioning}
\usetikzlibrary{shapes}
\usetikzlibrary{intersections}
\usetikzlibrary{arrows.meta}
\tikzset{font=\small}

\usepackage{csvsimple}
\usepackage{changepage}
\usepackage[tbtags]{mathtools}
\usepackage{siunitx}

\usepackage{csquotes}
\usepackage{glossaries}

\usetikzlibrary{external}
\usepackage{multirow}
\usepackage{tabularx}
\usepackage{booktabs}
\usepackage{array}
\usepackage{longtable}
\newcolumntype{L}[1]{>{\raggedright\let\newline\\\arraybackslash\hspace{0pt}}m{#1}}
\newcolumntype{C}[1]{>{\centering\let\newline\\\arraybackslash\hspace{0pt}}m{#1}}
\newcolumntype{R}[1]{>{\raggedleft\let\newline\\\arraybackslash\hspace{0pt}}m{#1}}

\DeclareSIUnit\pixel{px}

\definecolor{someprettyred}{rgb}{0.85, 0.0, 0.1}

\newacronym{ct}{CT}{computed tomography}
\newacronym{3d}{3D}{three-dimensional}
\newacronym{2d}{2D}{two-dimensional}

\begin{document}

\definecolor{brickred}{rgb}{0.8, 0.25, 0.33}
\definecolor{darkorange}{rgb}{1.0, 0.55, 0.0}
\definecolor{persiangreen}{rgb}{0.0, 0.65, 0.58}
\definecolor{persianindigo}{rgb}{0.2, 0.07, 0.48}
\definecolor{cadet}{rgb}{0.33, 0.41, 0.47}
\definecolor{turquoisegreen}{rgb}{0.63, 0.84, 0.71}
\definecolor{sandybrown}{rgb}{0.96, 0.64, 0.38}
\definecolor{blueblue}{rgb}{0.0, 0.2, 0.6}
\definecolor{ballblue}{rgb}{0.13, 0.67, 0.8}
\definecolor{greengreen}{rgb}{0.0, 0.5, 0.0} 

\title{Effect of Particle Size on the Suction Mechanism in Granular Grippers}
\subtitle{}

% Author Orchid ID: enter ID or remove command
\newcommand{\orcidauthorA}{0000-0002-7898-2986} % Add \orcidA{} behind the author's name
\newcommand{\orcidauthorB}{0000-0002-7218-4596} % Add \orcidB{} behind the author's name

\author{Angel Santarossa  \and
        Olfa D'Angelo \and
        Achim Sack \and
        Thorsten P\"oschel 
}


\institute{Angel Santarossa \and Olfa D'Angelo  \and  Achim Sack \and Thorsten P\"oschel \at
Institute for Multiscale Simulations\\
Friedrich-Alexander-Universit\"at Erlangen-N\"urnberg\\
Cauerstra\ss{}e 3, 91058 Erlangen\\
Germany\\
\email{thorsten.poeschel@fau.de}           %  \\
}

\date{Received: date / Accepted: date}

\maketitle

\begin{abstract} 
Granular grippers are highly adaptable end-effectors that exploit the reversible jamming transition of granular materials to hold and manipulate objects. Their holding force comes from the combination of three mechanisms: frictional forces, geometrical constraints, and suction effects. In this work, we experimentally study the effect of particle size on the suction mechanism. Through X-ray computed tomography, we show that small particles (average diameter $d \approx \SI{120}{\micro\meter}$) achieve higher conformation around the object than larger particles ($d \approx \SI{4}{\milli\meter}$), thus allowing the formation of air-tight seals. When the gripper is pulled off, mimicking lifting of an object, vacuum pressure is generated in the sealed cavity at the interface gripper--object. If the particles used as filling material are too large, the gripper does not conform closely around the object, leaving gaps between the gripper's membrane and the object. These gaps prevent the formation of sealed vacuum cavities between the object and the gripper and in turn hinder the suction mechanism from operating.

\end{abstract}

\section{Introduction}
\label{sec:intro}

Holding, manipulating, and transporting diverse objects are essential tasks in robotics.
The increasingly widespread use of robotics requires optimizing the efficiency, safety, and cost-effectiveness of such systems. 
More universal, i.e.~more versatile grippers are desirable,
which can hold objects of different shape, size, and surface properties. 
Rigid robotic grippers are only suitable for the manipulation of a limited range of objects. 
To increase their robustness and widen the range of objects that can be handled, 
these rigid grippers can incorporate sensory feedback and complex mechanics, 
for example grippers with multiple fingers~\cite{monkman_robot_2006,mason_robot_1985}, 
which mimic human hands. Human hand inspired systems require the use of complex control loops and multiple sensory input
to manipulate objects reliably.

A novel approach in the field of soft robotics
was introduced by Brown et al.~\cite{Brown2010universal},
who proposed to exploit the jamming transition in granular matter to grasp and hold a variety of objects~\cite{Majmudar2007, Behringer2018}. 
In its simplest design, a granular gripper is composed of a granulate contained in a flexible, non-permeable, and non-porous membrane. 
In this state, the granular material can flow when deformed, akin to a fluid. When the gripper is pressed against an object, the membrane and the granulate 
deform to conform to the object.
Then, to grip the object, the air is evacuated, causing the gripper's membrane to contract and compress the particles.
The granular material, enclosed in the membrane now under vacuum, 
assumes a static, solid-like state, 
characterized by a finite elastic modulus: the jammed state~\cite{liu2001jamming}. 
%in which its stiffness is significantly increased. 
When the granular material is compressed by the ambient pressure, the gripper pinches the target object, causing forces sufficient to grasp and lift it.
One remarkable feature of granular jamming is its reversibility: 
if air is allowed to flow back to the gripper,
the membrane relaxes, 
and the granulate returns to its fluid-like state, 
thus releasing the grasped object.
The granular gripper technology has proven to be highly adaptable, as it allows to hold a wide range of objects, including fragile ones, and even manipulate multiple objects simultaneously.~\cite{Brown2010universal, amendPositivePressureUniversal2012, amendSoftRoboticsCommercialization2016, licht_partially_2018, Kapadia2012design}.

As already described by Brown et al.~\cite{Brown2010universal}, 
three mechanisms contribute to the holding force of the granular gripper: 
friction, interlocking, and suction. 
Frictional forces are due to tangential stress at the contact of the membrane with the surface of the object. 
Interlocking occurs as the membrane wraps the object (or any protrusion of it) to the extent that when the granulate becomes rigid, geometrical restraints are produced between the gripper and the object. 
The suction mechanism is activated when sealed cavities appear between the membrane and the target object.
Interlocking was shown to be the most effective mechanism~\cite{Brown2010universal}.
However, its contribution highly depends on the shape of the object and 
the physical properties of the granular material,
which should allow the latter to flow and 
conform around the object to produce geometric constraints once jamming is induced.  
Because of the strength of interlocking and the universal presence of friction, 
research efforts have been directed toward their optimization~\cite{gotz_soft_2022, Kapadia2012design, licht_partially_2018}. 
Suction has hitherto attracted less attention, 
perhaps since 
the surface of the object has to be either smooth or wet
to produce a proper seal. 
Nonetheless, the high strength of suction is essential for holding fragile objects
or objects with a low static friction coefficient. 

In earlier work, it was assumed that the 
maximum holding force $F_{h}$ achieved by a granular gripper
is only marginally related to the physical properties of the granular material,
as long as it does not affect the degree of conformation of the gripper around the object~\cite{Brown2010universal,gomez-paccapeloEffectGranularMaterial2020} 
Recently, G\'{o}mez-Paccapelo et al.~\cite{gomez-paccapeloEffectGranularMaterial2020}
measured the friction force of the gripper for different granular materials. 
They found, in agreement with Brown et al.~\cite{Brown2010universal}, 
that the particle material has a minor effect on $F_{h}$ if the same penetration depth of the object into the gripper is achieved.

In the present work, 
we show that particle size affects the holding force produced through suction.  
Using X-ray computed tomography (CT), we reveal the mechanism behind this observation:  
particle size influences the formation of sealed cavities between the membrane and the 
object, therefore, the activation of the suction mechanism in granular grippers.

\section{Materials and methods}\label{secmethods}

\subsection{Experimental setup and measurement procedure}\label{sec1_2_methods}

\begin{figure*}[h]
\centering
\input{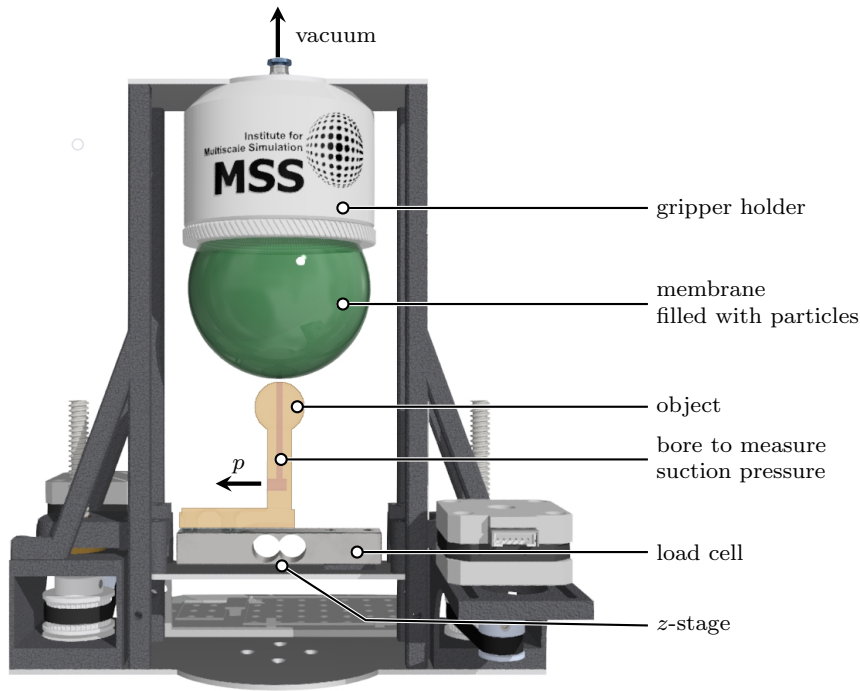}
\caption{Experimental setup: the granular gripper consists of a membrane filled with granular material, which grasps an object (here a sphere) placed underneath. The object, placed on a $z$-stage, can move vertically and is attached to a load cell to record the holding force,
and a bore on the top surface of the sphere allows to record the pressure $p$ at the membrane--object interface.
\label{fig:setup}}
\end{figure*}

Figure~\ref{fig:setup} shows a schematic of the experimental setup. The gripper consists of an elastic, air-tight, non-porous membrane of spherical shape of diameter 
 $d$ = $\SI{73.0}{\mm} \pm \SI{0.5}{\mm}$, 
filled with granular material. 
As granulate, we use two types of glass beads:
large ones, of average particle diameter $d_l$ = $\SI{4.0}{\mm} \pm \SI{0.3}{\mm}$,
and small ones, of average particle diameter $d_s$ = $\SI{120}{\micro\meter} \pm \SI{10}{\micro\meter}$.
The membrane is attached to a gripper holder. The gripper can be loaded with positive or negative air pressure. Underneath the gripper, the target object is placed on a platform that moves vertically, 
simulating the lifting of an object by the gripper. 
The object gripped is a plastic sphere of diameter $\SI{19.98}{\mm} \pm \SI{0.01}{\mm}$ whose surface was coated
using commercial lacquer
to be smooth and favor suction
at the interface between the membrane and the object. 
To measure the pressure at the interface membrane-object, the object has a bore hole at the top, which is connected to a pressure sensor.
To avoid suction, the sensor can be removed, preventing the formation of sealed cavities at the membrane-object interface.
The base on which the object is attached is displaced vertically utilizing two stepper motors. A force sensor is attached between the object and the platform to record the force applied to the object throughout the gripping process. 

Before each measurement, positive air pressure is applied to fluidize the granular material and 
erase the memory of previous gripping cycles.
A single measurement cycle comprises the following steps: 
(1)~The inner pressure of the gripper is balanced with the atmospheric pressure. 
(2)~The $z$-stage is elevated until reaching a fixed indentation depth. 
During this step, the object is pressed against the gripper while the latter conforms around the object. 
(3)~After a relaxation phase, the air is evacuated, causing the membrane to contract and the granulate to compact into its rigid, jammed state. 
The pressure difference created between the interior of the gripper's membrane and the atmosphere is $P_{vac} = \SI{90}{\kilo\pascal}$.
%the maximum achievable by our vacuum-pump system. 
(4)~The $z$-stage is moved down until the object detaches completely from the gripper.
Throughout the measurement cycle, the force acting on the object and
the pressure difference at the interface object--membrane is measured.

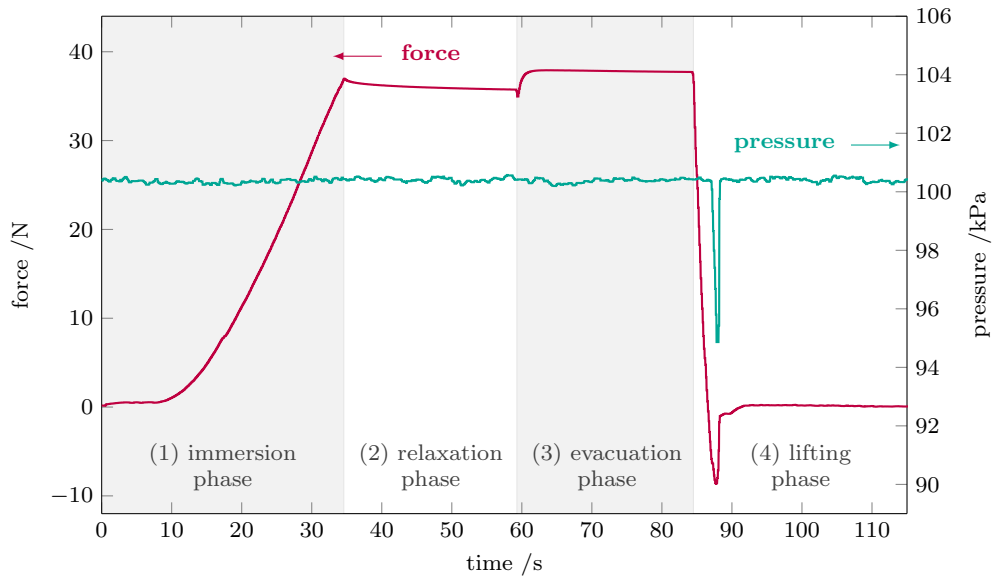
\begin{figure*}[h]
\centering
\begin{tikzpicture}
% let both axes use the same layers
\pgfplotsset{set layers}
\def\tone{34.57}
\def\ttwo{59.3}
\def\tthree{84.5}
\def\tfour{115}

\begin{axis}[
%scale only axis,
xmin=0, xmax=\tfour,
ymin=-12, ymax=44,
axis y line*=left,
xlabel=$x$,
ylabel style = {align=center},
ylabel={Plot 1 \ref{pgfplots:plot1}},
width=0.7\textwidth, height=0.47\textwidth,
%    enlarge x limits = 0.1,
%    legend style={at={(0.5,-0.38)},
%    anchor=north,legend columns=-1},
%    ylabel style={align=center, text width=2cm},
    ylabel={force~/\si{\newton}},
    xlabel={time~/\si{\second }}
    ]

\addplot 
[const plot, fill=gray, opacity=0.1, area legend, 
forget plot
]
coordinates
{(-20,-20)(\tone,-20)(\tone, 60)(-20, 60)}\closedcycle;

\addplot 
[const plot, fill=gray, opacity=0.1, area legend, 
forget plot
]
coordinates
{(\ttwo,-20)(\tthree,-20)(\tthree, 60)(\ttwo, 60)}\closedcycle;

\node at (0.5*\tone,-6.7) [text width=2cm, text centered, darkgray] {(1)~immersion\\phase};
\node at (47,-6.7) [text width=2cm, text centered, darkgray] {(2)~relaxation\\phase};
\node at (72.15,-6.7) [text width=2cm, text centered, darkgray] {(3)~evacuation\\phase};
\node at (99.8,-6.7) [text width=2cm, text centered, darkgray] {(4)~lifting\\phase};

\begin{scope}[xshift=0pt]
\node at (47,40) [text width=2cm, text centered, purple] (force) {\textbf{force}};
\draw[<-, >=latex, purple] (33,39.5) -- (40,39.5);
\end{scope}

\node at (97.5, 29.5) [text width=2cm, text centered, persiangreen] (pressure) {\textbf{pressure}};
\draw[->, >=latex, persiangreen] (107,29.5) -- (114,29.5);

\addplot
[  	thick, purple,
    no marks,
    line legend,
    enlarge x limits=false,
    ]
table   [
    x expr = \thisrow{time_s}, 
    y expr = \thisrow{force_Pa},
    col sep=tab
        ]
{force-signal-raw-data/force2.txt};

%\node at (rel axis cs: .98,.91) [left] {\textbf{(a)}};

\end{axis}

\begin{axis}[
%scale only axis,
axis y line*=right,
axis x line=none,
xmin=0, xmax=\tfour,
ymin=89, ymax=106,
ylabel style = {align=center},
width=0.7\textwidth, height=0.47\textwidth,
%    enlarge x limits = 0.1,
%    legend style={at={(0.5,-0.38)},
%    anchor=north,legend columns=-1},
%    ylabel style={align=center, text width=2cm},
    ylabel={pressure~/\si{\kilo\pascal}},
]

\addplot
[  	no marks, thick, persiangreen,
    line legend,
    enlarge x limits=false,
    ]
table   [
    x expr = \thisrow{time_s}, 
    y expr = \thisrow{pressure_kPa},
    col sep=tab
        ]
{force-signal-raw-data/pressure2.txt};

\end{axis}

\end{tikzpicture}
\caption{Force in the $z$-direction (red curve, left ordinate axis) and pressure at the object--membrane interface (blue curve, right ordinate axis) as a function of time. The granular gripper is filled by small glass beads. The data shown correspond to a full gripping cycle, steps (1) to (4).
\label{fig:force-signal}}
\end{figure*}

Figure~\ref{fig:force-signal} shows both force and pressure signals during a single measurement for the small glass beads.
The different phases of the gripping process are specified: first, force in the $z$-direction increases, corresponding to the object being pushed against the gripper. 
During the relaxation phase, a small decrease in the force is caused by the sudden stop of the $z$-stage motion, followed by a transition to a constant value, produced by the reorganization of the granulate. 
Subsequently, during the evacuation phase, the force increases slightly and reaches a constant value. 
Afterwards, the $z$-stage is moved down to pull the object from the gripper. The force assumes negative values, which is the holding force acting on the object. When the object detaches from the gripper, 
then the measured force relaxes to zero. 
The absolute value of the minimum force corresponds 
to the maximum holding force attained by the gripper, $F_{h}$.

The pressure in the gap fluctuates around atmospheric pressure 
until the $z$-stage is moved down. 
The pressure signal then decreases, reaching a minimum value at the beginning of the lifting phase.
When the seal around the object is broken, it increases again to atmospheric pressure. 
This decrease in pressure corresponds to 
the instant at which the suction mechanism becomes active:
the force due to suction is not yielded during the evacuation phase, 
but rather when the object is pulled away from the gripper. 
As the object is moved down, sealed cavities form and increase in volume, 
by decreasing the pressure inside the cavities,
hence producing suction.
As the object continues moving downwards, 
the seal eventually breaks 
and pressure increases to equalize with environmental pressure.

\subsection{Interior imaging: X-ray computed tomography}

The force exerted by the gripper on the object studied
so far is due to the collective motion of the granular particles. 
In the current section, we consider
the motion of the grains in the curse of the
gripping process. 
To this end, we determine the location of each
particle by means of X-ray CT.
The experimental setup presented in Fig.~\ref{fig:setup}
is made from low X.ray absorbent materials,
making it suitable for use in an X-ray tomograph.

The X-ray tomographies are recorded 
in a large laboratory tomograph containing the entire experimental setup.
Parameters used for acquisition are given in Table~\ref{tab:parameters_ct}.
A \SI{1.5}{\mm} copper plate is used to filter the X-ray, 
which significantly reduces beam hardening effects~\cite{baur_correction_2019}. 
%A resolution of 41.5 px/mm is achieved. 
%The obtained radiographs are then employed to obtain a three-dimensional reconstruction of the granular system. To this, we use 
The software XrayOffice (v2.0)
is used to reconstruct a \gls{3d} representation of the granular packing.

\begin{table*}[]
\centering
\small 
\caption{\label{tab:parameters_ct}
Parameters used for X-ray CT.
}
\renewcommand{\arraystretch}{1.1} 
\begin{tabular}[c]{C{0.15\textwidth}C{0.08\textwidth}C{0.08\textwidth}C{0.11\textwidth}C{0.13\textwidth}C{0.1\textwidth}C{0.12\textwidth}}
\toprule
		\textbf{Detector}	&	\textbf{Source voltage}	&	 \textbf{Target current}	
		& \textbf{Projections per scan}	& \textbf{Measurements per projection }	& \textbf{Exposure time}		& \textbf{Resolution}	\\
\midrule
		DEXELA 1512 14 bit flat panel (see~\cite{menendez_x-ray_2019}) & \SI{150}{\kilo\volt} &	\SI{260}{\micro\ampere}	&	 1600  & 10	& \SI{100}{\milli\second}	& {85.8} \si{\pixel\per\micro\m} 	\\
\bottomrule
\end{tabular}
\end{table*}

The positions of the particles are detected by image analysis
using the Image Processing Toolbox (Matlab~2019) and 
the \texttt{volume2positions} package~\cite{weis_analyzing_2017}.
We apply the following algorithm:
(1) A bilateral filter is applied to 
reduce noise while conserving the edges of the elements in the image.
It modifies the intensity of each voxel by a weighted average of the intensity value from nearby voxels.
(2) The image is binarized to separate particles from the background.
(3) The Euclidean distance map and the watershed algorithm are applied to assign a particle ID to each particle \cite{weis_analyzing_2017}.
(4) By computing the centroid of all its voxels, the position of each particle is obtained. 
A full \gls{3d} of the particle system containing the corresponding position of each particle is obtained. 

\section{Results and discussions}\label{sec2_results}

To determine the contribution of suction to the total force,
we measure the maximum holding force achieved by the granular gripper  
in two different configurations, labeled \emph{closed} and \emph{opened}. 
In the closed configuration, the bore hole in the object (see Fig.~\ref{fig:setup}) is sealed by a pressure sensor, which allows a differential pressure to be created at the membrane--object interface. On the contrary, in the opened configuration,
the bore hole is not sealed, that is, absent vacuum can not cause suction.
We perform experiments with both configurations for an object with a wetted surface, to aid the formation of an air-tight seal. 
Besides, those experiments are conducted with a gripper filled by two different glass bead sizes:
average diameters of \SI{4.0}{\mm} (large particles) and \SI{120}{\micro\meter} (small particles).

\begin{figure*}[h]
\begin{tikzpicture}
\begin{axis}[
    width=0.5\textwidth, height=0.3\textwidth,
    ybar, %xtick=data,
    ymin=0,ymax=6.9,
%    enlarge x limits = 0.1,
%    legend style={at={(0.5,-0.38)},
%    anchor=north,legend columns=-1},
%    ylabel style={align=center, text width=2cm},
    ylabel={holding force~/\si{\newton}},
%    symbolic x coords={wet-open, wet-close, dry-open, dry-close},
	xmin=0.5,xmax=4.5,
	xtick={1,2,3,4},
	x tick style={draw=none},
	xticklabels={wet-open, wet-close, dry-open, dry-close},
	xticklabel style = {font=\footnotesize},
%    nodes near coords, 
	every node near coord/.style={
        opacity=1, 
        text depth=6.5mm,
        /pgf/number format/precision=2
        },
	every axis plot/.append style={
          ybar,
          bar width=.5,
          bar shift=0pt,
          fill
        }
    ]

\addplot [fill = blue!50, font=\scriptsize, opacity=0.75,
        error bars/.cd,
        y dir=both,
        y explicit relative, 
        error bar style={color=black, solid, opacity=1}
        ] 
coordinates {
    (1, 2.91) +- (0.41,0.41)
    };

\addplot [fill = blue, font=\scriptsize, opacity=0.75,
        error bars/.cd,
        y dir=both,
        y explicit relative, 
        error bar style={color=black, solid, opacity=1}
        ] 
coordinates {
    (2, 2.894) +- (0.515,0.515)
    };
\addplot [fill = orange!50, font=\scriptsize, opacity=0.75,
        error bars/.cd,
        y dir=both,
        y explicit relative, 
        error bar style={color=black, solid, opacity=1}
        ] 
coordinates {
    (3, 2.78) +- (0.0294,0.294)
	};
        
\addplot [fill = orange, font=\scriptsize, opacity=0.75,
        error bars/.cd,
        y dir=both,
        y explicit relative, 
        error bar style={color=black, solid, opacity=1}
        ] 
coordinates {    
	(4, 3.23) +- (0.257,0.257) 
    };

\node at (rel axis cs: .98,.91) [left] {\textbf{(a)}};

\node at (rel axis cs: .5,.91) [] {$d=\SI{4.0}{\mm}$};

\end{axis}

\end{tikzpicture}\quad\quad
\begin{tikzpicture}
\begin{axis}[
    width=0.5\textwidth, height=0.3\textwidth,
    ybar, %xtick=data,
    ymin=0,ymax=600,
%    enlarge x limits = 0.1,
%    legend style={at={(0.5,-0.38)},
%    anchor=north,legend columns=-1},
%    ylabel style={align=center, text width=2cm},
    ylabel={pressure~/\si{\pascal}},
%    symbolic x coords={wet-open, wet-close, dry-open, dry-close},
	xmin=0.5,xmax=4.5,
	xtick={1,2,3,4},
	x tick style={draw=none},
	xticklabels={wet-open, wet-close, dry-open, dry-close},
	xticklabel style = {font=\footnotesize},
%    nodes near coords, 
	every node near coord/.style={
        opacity=1, 
        text depth=3.5mm,
        /pgf/number format/precision=0
        },
	every axis plot/.append style={
          ybar,
          bar width=.5,
          bar shift=0pt,
          fill
        }
    ]

\addplot [fill = blue!50, font=\scriptsize, opacity=0.75,
        error bars/.cd,
        y dir=both,
        y explicit relative, 
        error bar style={color=black, solid, opacity=1}
        ] 
coordinates {
    (1, 143.30) +- (0.48,0.48)
    };

\addplot [fill = blue, font=\scriptsize, opacity=0.75,
        error bars/.cd,
        y dir=both,
        y explicit relative, 
        error bar style={color=black, solid, opacity=1}
        ] 
coordinates {
    (2, 100.47) +- (0.32,0.32)
    };
\addplot [fill = orange!50, font=\scriptsize, opacity=0.75,
        error bars/.cd,
        y dir=both,
        y explicit relative, 
        error bar style={color=black, solid, opacity=1}
        ] 
coordinates {
    (3, 141.87) +- (0.57,0.57)
	};
        
\addplot [fill = orange, font=\scriptsize, opacity=0.75,
        error bars/.cd,
        y dir=both,
        y explicit relative, 
        error bar style={color=black, solid, opacity=1}
        ] 
coordinates {    
	(4, 247.8) +- (0.35,0.35) 
    };

\node at (rel axis cs: .98,.91) [left] {\textbf{(b)}};

\node at (rel axis cs: .5,.91) [] {$d=\SI{4.0}{\mm}$};

\end{axis}

\end{tikzpicture}\\
\begin{tikzpicture}
\begin{axis}[
    width=0.5\textwidth, height=0.3\textwidth,
    ybar, %xtick=data,
    ymin=0,ymax=8.2,
%    enlarge x limits = 0.1,
%    legend style={at={(0.5,-0.38)},
%    anchor=north,legend columns=-1},
%    ylabel style={align=center, text width=2cm},
    ylabel={holding force~/\si{\newton}},
%    symbolic x coords={wet-open, wet-close, dry-open, dry-close},
	xmin=0.5,xmax=4.5,
	xtick={1,2,3,4},
	x tick style={draw=none},
	xticklabels={wet-open, wet-close, dry-open, dry-close},
	xticklabel style = {font=\footnotesize},
%    nodes near coords, 
	every node near coord/.style={
        opacity=1, 
        text depth=2mm,
        /pgf/number format/precision=2
        },
	every axis plot/.append style={
          ybar,
          bar width=.5,
          bar shift=0pt,
          fill
        }
    ]

\addplot [fill = blue!50, font=\scriptsize, opacity=0.75,
        error bars/.cd,
        y dir=both,
        y explicit relative, 
        error bar style={color=black, solid, opacity=1}
        ] 
coordinates {
    (1, 3.62) +- (0.07,0.07)
    };

\addplot [fill = blue, font=\scriptsize, opacity=0.75,
        error bars/.cd,
        y dir=both,
        y explicit relative, 
        error bar style={color=black, solid, opacity=1}
        ] 
coordinates {
    (2, 6.29) +- (0.064,0.064)
    };
\addplot [fill = orange!50, font=\scriptsize, opacity=0.75,
        error bars/.cd,
        y dir=both,
        y explicit relative, 
        error bar style={color=black, solid, opacity=1}
        ] 
coordinates {
    (3, 4.05) +- (0.099,0.099)
	};
        
\addplot [fill = orange, font=\scriptsize, opacity=0.75,
        error bars/.cd,
        y dir=both,
        y explicit relative, 
        error bar style={color=black, solid, opacity=1}
        ] 
coordinates {    
	(4, 3.91) +- (0.036,0.036) 
    };

\node at (rel axis cs: .98,.91) [left] {\textbf{(c)}};

\node at (rel axis cs: .5,.91) [] {$d=\SI{120}{\micro\meter}$};

\end{axis}

\end{tikzpicture}\quad\quad
\begin{tikzpicture}
\begin{axis}[
    width=0.5\textwidth, height=0.3\textwidth,
    ybar, %xtick=data,
%    ymode=log,
    ymin=0,ymax=10000,
%    enlarge x limits = 0.1,
%    legend style={at={(0.5,-0.38)},
%    anchor=north,legend columns=-1},
%    ylabel style={align=center, text width=2cm},
    ylabel={pressure~/\si{\pascal}},
%    symbolic x coords={wet-open, wet-close, dry-open, dry-close},
	xmin=0.5,xmax=4.5,
	xtick={1,2,3,4},
	x tick style={draw=none},
	xticklabels={wet-open, wet-close, dry-open, dry-close},
	xticklabel style = {font=\footnotesize},
%    nodes near coords, 
	every node near coord/.style={
		/pgf/number format/1000 sep=,
        opacity=1, 
        text depth=3.5mm,
        /pgf/number format/precision=0,
        },
	every axis plot/.append style={
          ybar,
          bar width=.5,
          bar shift=0pt,
          fill
        }
    ]

\addplot [fill = blue!50, font=\scriptsize, opacity=0.75,
        error bars/.cd,
        y dir=both,
        y explicit relative, 
        error bar style={color=black, solid, opacity=1}
        ] 
coordinates {
    (1, 315) +- (0.22,0.22)
    };

\addplot [fill = blue, font=\scriptsize, opacity=0.75,
        error bars/.cd,
        y dir=both,
        y explicit relative, 
        error bar style={color=black, solid, opacity=1}
        ] 
coordinates {
    (2, 6415.0) +- (0.178,0.178)
    };
\addplot [fill = orange!50, font=\scriptsize, opacity=0.75,
        error bars/.cd,
        y dir=both,
        y explicit relative, 
        error bar style={color=black, solid, opacity=1}
        ] 
coordinates {
    (3, 236) +- (0.258,0.258)
	};
        
\addplot [fill = orange, font=\scriptsize, opacity=0.75,
        error bars/.cd,
        y dir=both,
        y explicit relative, 
        error bar style={color=black, solid, opacity=1}
        ] 
coordinates {    
	(4, 149) +- (0.464,0.464) 
    };

\node at (rel axis cs: .98,.91) [left] {\textbf{(d)}};

\node at (rel axis cs: .5,.91) [] {$d=\SI{120}{\micro\meter}$};

\end{axis}

\end{tikzpicture}
\centering
\caption{Maximum holding force~(a, c) and maximum difference between ambient pressure and pressure within the gap (b, d),
obtained for grasping a smooth sphere with the gripper filled by large glass beads (average diameter \SI{4.0}{\mm}) (a, b),
and small glass beads (average diameter \SI{120}{\micro\meter}) (c, d).
Data is shown for a combination of {opened}/{closed} and wet/dry conditions.
Each result is averaged over six independent measurements; error bars correspond to the standard deviation of the mean. 
\label{fig:holdingF+pressure}}
\end{figure*}
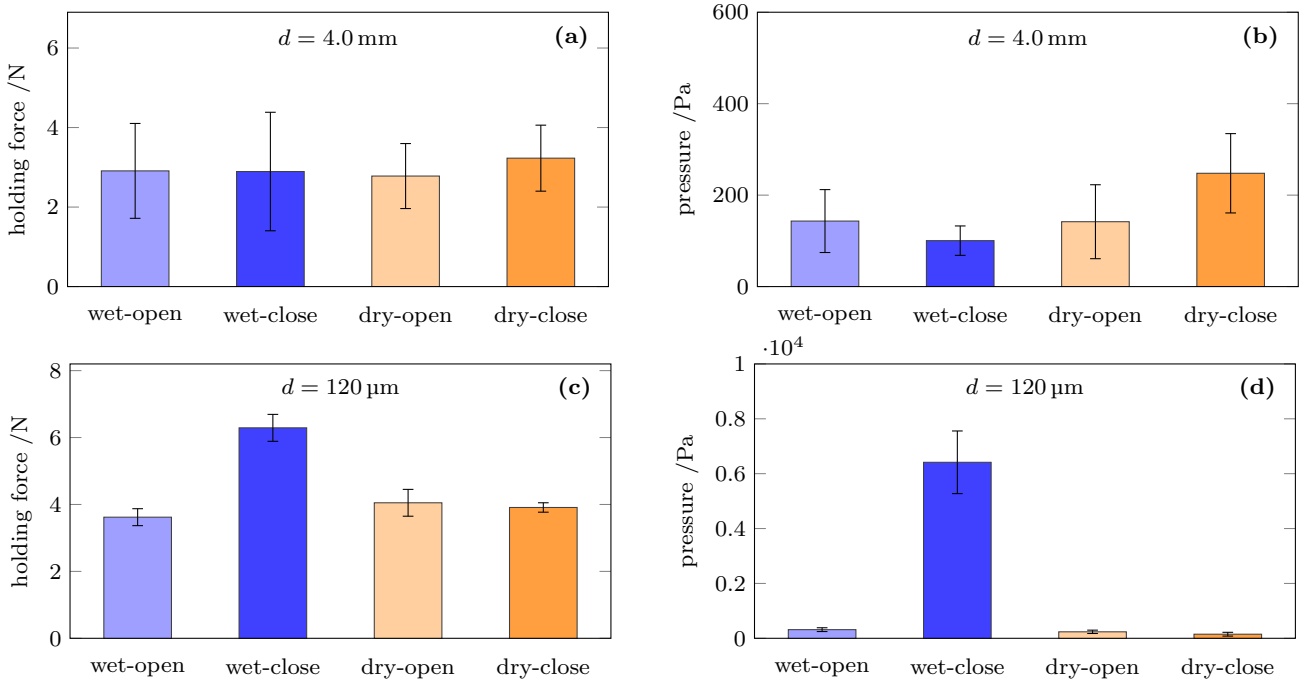

Figure~\ref{fig:holdingF+pressure} shows the maximum holding force and maximum differential pressure, i.e, the difference between ambient pressure and
pressure measured at the interface membrane--object, 
for both types of particles.
For large particles (Fig.~\ref{fig:holdingF+pressure}(a, c)),
the holding force is similar
regardless of the configuration (opened or closed and wet or dry).  
The low differential pressure measured in all cases (Fig.~\ref{fig:holdingF+pressure}(b)) indicates that an airtight seal is not formed during the gripping experiments; therefore,
the contribution of suction to the gripping force is negligible.

For the gripper filled with small particles,
for a wet surface of the object, the holding force is increased in the closed configuration 
compared to the opened configuration (Fig.~\ref{fig:holdingF+pressure}(c)),
indicating a significant contribution of suction to the holding force. 
This agrees with a noticeable increase in the differential pressure (Fig.~\ref{fig:holdingF+pressure}(d)). 
Using $F_{s}=P_{g}A$, where $F_{s}$ is the force due to suction, $P_{g}$ is the difference between ambient pressure and the pressure within the gap, and $A$ the horizontal cross-section area, we estimate the maximum suction force that can be achieved. $A$ is limited by the diameter of the spherical object. 
With the measured value of $P_{g}$, 
we obtain $F_{s} \approx \SI{2.02}{\newton}$,
which agrees with the difference between the
opened and closed configurations.
In contrast, if the object's surface is dry,
there is no appreciable difference in the holding force
for \emph{opened} and \emph{closed} configurations.
We conclude that there is no significant difference in the pressure inside the gap (Fig.~\ref{fig:holdingF+pressure}(b)) because no airtight cavities are formed.

\begin{figure*}
\centering
\input{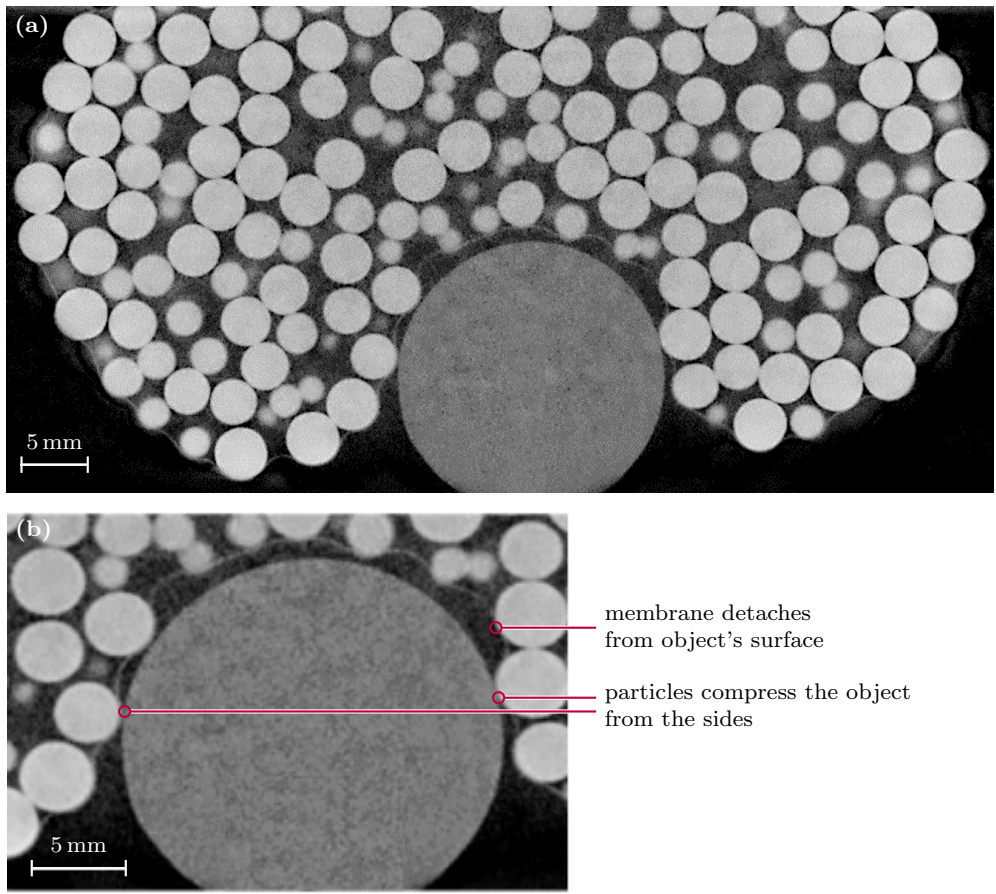}
\caption{X-ray tomogram slice for large particles, after evacuation of the air from the gripper: original snapshot (a) and magnification in the region around the object (b).
\label{fig:slice+displacement}}
\end{figure*} 

The force as an integral value can be understood
if we consider the granular packing at different stages of the experiment. 
To this end,
we record tomograms at two stages of the experiment:
(1) when the gripper has conformed around the object, but before evacuating the air within the membrane; (2) after jamming of the packing.
Figure~\ref{fig:slice+displacement} shows a cut through the center of the gripper filled by large particles after evacuation.
Particles are in contact with the object on each side.
Above the object, the membrane detaches from the object's surface, forming cavities at the 
membrane--object interface. We believe that such
cavities are the basis of the suction mechanism:
as the object is lifted, the cavities try to expand due to the weight of the object.
If the cavities are sealed, the pressure lowers inside of them,
producing a force to maintain their volume constant.
This counter force is the contribution to the holding force 
due to suction, $F_s$.

\begin{figure*}
\centering
\input{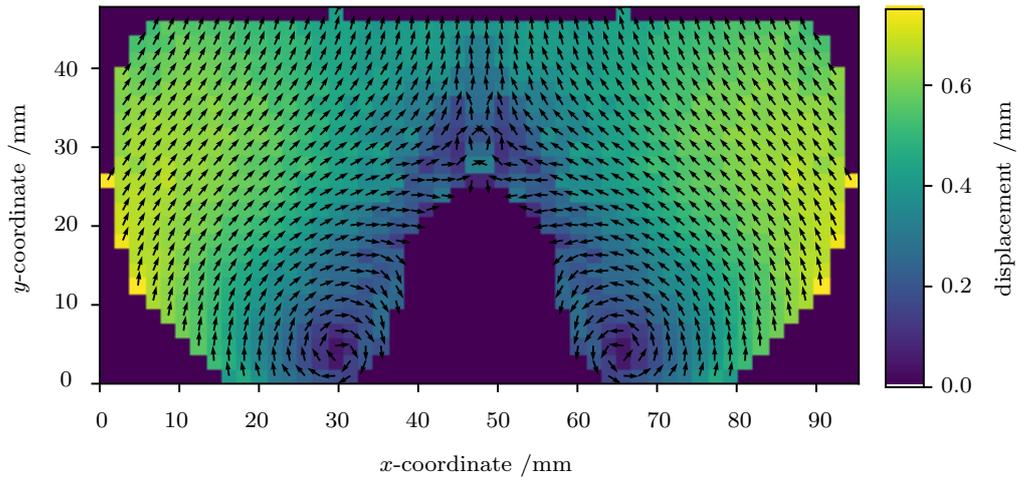}
\caption{~Displacement field of the particles within the gripper, obtained from tomograms taken before and after vacuum is applied.
\label{fig:displacement}}
\end{figure*}

Figure~\ref{fig:slice+displacement} is a \gls{2d} snapshot of the system.
To analyze the dynamic behavior of the granulate in \gls{3d} we compute the particles' displacement field,
between the end of the relaxation phase (step(2) in figure~\ref{fig:force-signal}) and the end of 
the evacuation phase (step(3) in figure~\ref{fig:force-signal}).
The displacement field 
for the large particles is shown in figure~\ref{fig:displacement}. 
Using cylindrical coordinates, the displacement field is obtained by azimuthal space average.
The image is composed of twice the center right section.
The plot is mirrored for better visualization. 
The values of the pixels closer to the center of the image and the object 
might not be statistically significant due to average over few particle displacements. 
We assume similar displacement regardless of particles' size. 
The particles located on the upper sides of the object move towards it.
We infer that this horizontal displacement towards the object
produces the grasping, as the granulate conforms to the object's shape.
The particles above the object displace upward, which creates the cavities observed in figure~\ref{fig:slice+displacement} at the membrane--object interface.
As shown by the pressure peak during the lifting phase (see figure~\ref{fig:holdingF+pressure}),
if those cavities are sealed, the increase in the differential pressure within the cavities
produces suction and in turn a higher contribution to the maximum holding force.
Therefore, we hypothesize that the lower holding force measured for large particles, compared to small ones,
is because the gripper with large particles can not closely conform to the object. Hence, the cavities at the object--membrane interface are not sealed, and the suction mechanism is not activated.  

\begin{figure*}[h!]
\input{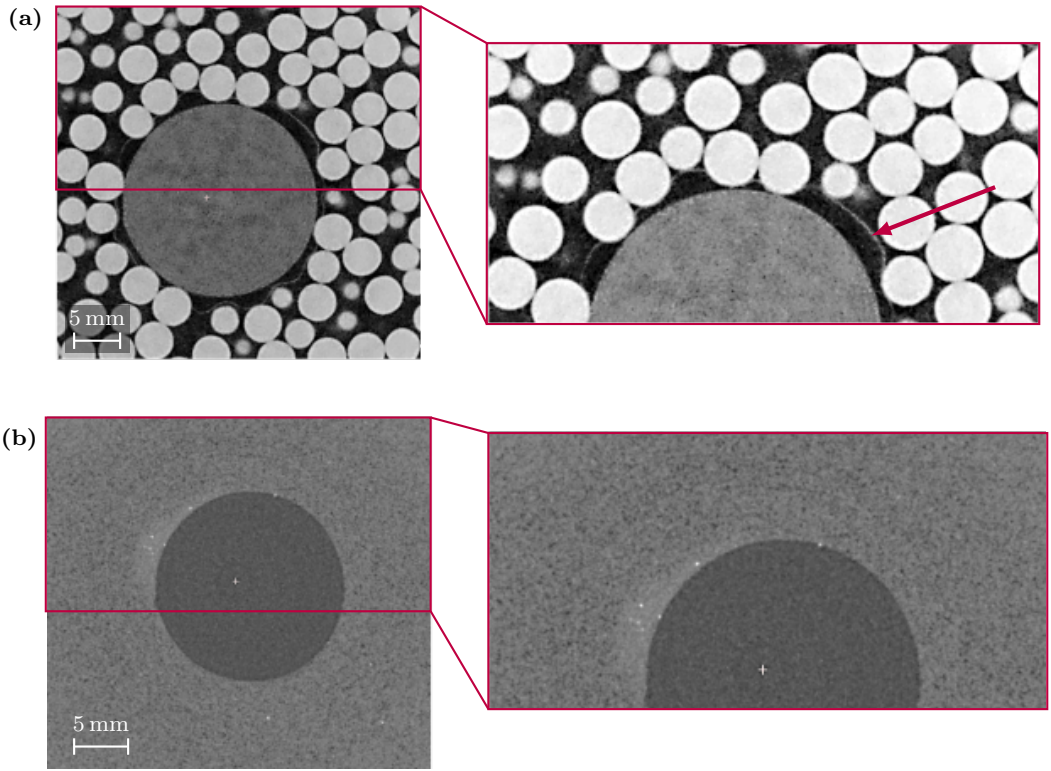}
\centering
\caption{Tomogram slices for (a) ($d_l$ = $\SI{4.0}{\milli\meter}$) particles and (b) ($d_s$ = $\SI{120}{\micro\meter}$) particles after evacuation of the gripper. The images on the right side are magnifications of the original tomogram slices. The arrow notes a region where the membrane is not in contact with the surface of the object.\label{fig:slice-size-comparison}}
\end{figure*}

Figure~\ref{fig:slice-size-comparison} shows a horizontal cut through the granular packing within the gripper, after evacuation,
for large and small particles (respectively Figs.~\ref{fig:slice-size-comparison}(a) and \ref{fig:slice-size-comparison}(b)). 
For large particles, the gripper does not conform closely around the object: 
large areas where the gripper's membrane is not in contact with the object are visible. 
If these cavities are not sealed, air will flow within these gaps as the object is pulled-down, and suction will not activate. 
For small particles, the gripper closely conforms to the object, therefore, 
gaps between the membrane and the object are not visible. 
Due to higher compliance of the gripper with small particles,
the cavities at the membrane--object interface seal, leading to the activation of the suction mechanism, 
as confirmed by the difference in holding force shown in Fig.~\ref{fig:holdingF+pressure}.
Holding forces are similar for small and large particles
if sealed cavities are not formed
at the object--membrane interface (gripper in opened configuration). 
For the object with a wetted surface in the \emph{closed} configuration, 
the contribution of suction to the holding force is due to small particles only.
These findings explain the effect of granulate size 
on the contribution of suction to the holding force of granular grippers:
large particles leave large gaps around the object between two particles.
On the contrary, a gripper filled with smaller particles, by closely conforming around the object,
can produce sealed cavities at the membrane--object interface, which is magnified
by wetting the object's surface. Such sealed cavities will produce
the counter-force due to suction when the object is lifted,
increasing significantly the maximum holding force.

\section{Summary and outlook}\label{sec3}

We experimentally study the effect of particle size on the suction mechanism in granular grippers. 
We measure the maximum holding force achieved 
by a granular gripper and the maximum difference between ambient pressure and the pressure at the interface object--membrane under different conditions: 
a combination of opened/closed configuration of the system and a dry/wet surface of the object.
In the \emph{opened} configuration, a borehole on the object is left open to prevent a pressure gradient at the membrane--object interface 
and impede the activation of suction mechanism. In the \emph{closed} configuration, the borehole on the object is sealed by connecting
a pressure sensor. The surface of the object is made wet to ease the sealing at the membrane--object interface. 

Through X-ray \acrlong{ct}, we link the activation of suction to the size of the filling particles.
When small particles (diameter $d_s$ = $\SI{120}{\micro\meter}$) 
are used, the gripper closely conforms around the object. 
Air-tight seals can form between the gripper and the object, 
and suction is activated. 
For large particles (diameter $d_l$ =$\SI{4.0}{\mm}$), 
the gripper's membrane does not fully conforms around the object. 
Gaps remain between the gripper's membrane an the object, which
let the pressure at the membrane--object interface equalize
with ambient pressure, hindering 
the formation of sealed cavities and, therefore,
impeding the suction mechanism to activate.

Our results can be applied to enhance granular gripping systems 
by making them more robust against
changes in geometry or surface properties of target objects. 
More generally, our findings can be applied to design 
granular jamming-based soft robots more efficiently 
by controlling the activation of the suction mechanism.

\begin{acknowledgements}
A.S. and T.P. gratefully acknowledge funding by Deutsche 
Forschungsgemeinschaft (DFG, German Research Foundation)--Project Number 
411517575. The authors thank Walter Pucheanu for his contribution to the design and construction of the experimental setup. This work was supported by the Interdisciplinary Center for Nanostructured Films (IZNF), the Competence Unit for Scientific Computing (CSC), and the Interdisciplinary Center for Functional Particle Systems (FPS) at Friedrich-Alexander-Universit\"at Erlangen-N\"urnberg.

\end{acknowledgements}

\section*{Compliance with ethical standards}
The authors declare that they have no conflict of interest.

% \bibliographystyle{spphys}       % APS-like style for physics

\bibliographystyle{ieeetr}

\bibliography{additionalBib.bib}   % name your BibTeX data base

\begin{thebibliography}{10}

\bibitem{monkman_robot_2006}
G.~J. Monkman, S.~Hesse, R.~Steinmann, and H.~Schunk, {\em Robot {Grippers}}.
\newblock Wiley, Oct. 2006.

\bibitem{mason_robot_1985}
M.~T. Mason and J.~K. Salisbury, {\em Robot Hands and the Mechanics of
  Manipulation}.
\newblock Cambridge, Mass: MIT Press, 1985.

\bibitem{Brown2010universal}
E.~Brown, N.~Rodenberg, J.~Amend, A.~Mozeika, E.~Steltz, M.~R. Zakin,
  H.~Lipson, and H.~M. Jaeger, ``Universal robotic gripper based on the jamming
  of granular material,'' {\em Proceedings of the National Academy of
  Sciences}, vol.~107, no.~44, pp.~18809--18814, 2010.

\bibitem{Majmudar2007}
T.~S. Majmudar, M.~Sperl, S.~Luding, and R.~P. Behringer, ``Jamming transition
  in granular systems,'' {\em Phys. Rev. Lett.}, vol.~98, p.~058001, 2007.

\bibitem{Behringer2018}
R.~P. Behringer and B.~Chakraborty, ``The physics of jamming for granular
  materials: A review,'' {\em Reports on Progress in Physics}, vol.~82,
  p.~012601, 2018.

\bibitem{liu2001jamming}
A.~J. Liu and S.~R. Nagel, eds., {\em Jamming and rheology: constrained
  dynamics on microscopic and macroscopic scales (1st ed.)}.
\newblock CRC Press, 2001.

\bibitem{amendPositivePressureUniversal2012}
J.~R. Amend, E.~Brown, N.~Rodenberg, H.~M. Jaeger, and H.~Lipson, ``A
  {{positive pressure universal gripper based}} on the {{jamming}} of
  {{granular material}},'' {\em IEEE Transactions on Robotics}, vol.~28,
  pp.~341--350, Apr. 2012.

\bibitem{amendSoftRoboticsCommercialization2016}
J.~Amend, N.~Cheng, S.~Fakhouri, and B.~Culley, ``Soft {{robotics
  commercialization}}: {{Jamming grippers}} from {{research}} to {{product}},''
  {\em Soft Robot}, vol.~3, pp.~213--222, Dec. 2016.

\bibitem{licht_partially_2018}
S.~Licht, E.~Collins, G.~Badlissi, and D.~Rizzo, ``A {Partially} {filled}
  {jamming} {gripper} for {underwater} {recovery} of {objects} {resting} on
  {soft} {surfaces},'' in {\em 2018 {IEEE}/{RSJ} {International} {Conference}
  on {Intelligent} {Robots} and {Systems} ({IROS})}, (Madrid), pp.~6461--6468,
  IEEE, Oct. 2018.

\bibitem{Kapadia2012design}
J.~Kapadia and M.~Yim, ``Design and performance of nubbed fluidizing jamming
  grippers,'' in {\em 2012 IEEE International Conference on Robotics and
  Automation}, pp.~5301--5306, 2012.

\bibitem{gotz_soft_2022}
H.~Götz, A.~Santarossa, A.~Sack, T.~Pöschel, and P.~Müller, ``Soft particles
  reinforce robotic grippers: robotic grippers based on granular jamming of
  soft particles,'' {\em Granular Matter}, vol.~24, p.~31, Feb. 2022.

\bibitem{gomez-paccapeloEffectGranularMaterial2020}
J.~M. {G{\'o}mez{\textendash}Paccapelo}, A.~A. Santarossa, H.~D. Bustos, and
  L.~A. Pugnaloni, ``Effect of the granular material on the maximum holding
  force of a granular gripper,'' {\em Granular Matter}, vol.~23, no.~1, p.~4,
  2020.

\bibitem{baur_correction_2019}
M.~Baur, N.~Uhlmann, T.~Pöschel, and M.~Schröter, ``Correction of beam
  hardening in {X}-ray radiograms,'' {\em Review of Scientific Instruments},
  vol.~90, p.~025108, Feb. 2019.

\bibitem{menendez_x-ray_2019}
H.~T. Menendez, M.~Heckel, A.~Sack, and T.~Pöschel, ``X-ray tomography in
  micro-gravity,'' {\em Review of Scientific Instruments}, vol.~90, p.~105103,
  Oct. 2019.

\bibitem{weis_analyzing_2017}
S.~Weis and M.~Schröter, ``Analyzing {X}-ray tomographies of granular
  packings,'' {\em Review of Scientific Instruments}, vol.~88, p.~051809, May
  2017.

\end{thebibliography}

\end{document}